\documentclass[twocolumn,prX,groupedaddress,showpacs,amsfonts,amssymb,amsmath,letterpaper]{revtex4-1}
\usepackage{amsmath}
\usepackage{graphicx}
\usepackage{amssymb}
\usepackage{epstopdf}
\usepackage{adjustbox}
\usepackage{mathrsfs}
\usepackage{algorithm}
\usepackage[noend]{algpseudocode}

\makeatletter

\newcommand{\Rmnum}[1]{\expandafter\@slowromancap\romannumeral #1@}
\makeatother
\usepackage{color}
\begin{document}

\title{Measurement-only topological quantum computation without forced measurements}

\author{Huaixiu Zheng}
\author{Arpit Dua}
\author{Liang Jiang}
\affiliation{\textit{Departments of Applied Physics and Physics, Yale University, New Haven, Connecticut, USA}}
\affiliation{\textit{Yale Quantum Institute, Yale University, New Haven, CT 06520, USA}}

\date{\today}

\begin{abstract}
We investigate the measurement-only topological quantum computation (MOTQC) approach proposed by Bonderson et al. [Phys. Rev. Lett. 101, 010501 (2008)] where the braiding operation is shown to be equivalent to a series of topological charge ``forced measurements'' of anyons.
In a forced measurement, the charge measurement is forced to yield the desired outcome (e.g.~charge 0) via repeatedly measuring charges in different bases.
This is a probabilistic process with a certain success probability for each trial. In practice, the number of measurements needed will vary from run to run.
We show that such an uncertainty associated with forced measurements can be removed by simulating the braiding operation using a fixed number of three measurements supplemented by a correction operator.
Furthermore, we demonstrate that in practice we can avoid applying the correction operator in hardware by implementing it in software.
Our findings greatly simplify the MOTQC proposal and only require the capability of performing charge measurements to implement topologically protected transformations generated by braiding exchanges without physically moving anyons.

\end{abstract}


\maketitle

\section{Introduction}
Anyons are excitations of topological phases and exhibit exotic exchange statistics \cite{NayakRMP08,SternSci13}, different from either fermions or bosons.
In particular, non-Abelian anyons which obey noncommutative exchange statistics can be used to encode and process quantum information in the associated topological degenerate subspace for 
topological quantum computation (TQC).
Such a subspace is characterized by topology, and is robust against local perturbations.
This intrinsic error-protection holds great promise for fault-tolerant topological quantum computation and has triggered a lot of interest in searching for non-Abelian anyons.

Examples of non-Abelian anyons with realistic proposals of physical setups include Majorana \cite{Kitaev01,LutchynPRL10,OregPRL10} and parafermion zero modes \cite{AliceaARCMP16}.
Several experiments have found convincing evidence of the existence of Majorana zero modes in systems of semiconductor nanowires in proximity to a superconductor \cite{MourikSci12,DasNatPhys12,RokhinsonNatPhys12,DengNL12,FinckPRL13} and  magnetic ad-atomic chains placed on the surface of a superconductor \cite{PergeSci14}.
Recently, the characteristic exponential energy splitting of Majorana zero modes has been observed in  proximity-induced superconducting
Coulomb islands \cite{AlbrechtNat16}.
Such an exponential protection implies that quantum information
can be encoded in the degenerate topological states in a nonlocal manner.
Rotations on the logical space can be performed through braiding exchanges of Majorana modes, which only depend on the topology of the braiding path and are robust against local noise \cite{NayakRMP08}.
However, the set of operations is limited to Clifford gates and is not complete for universal quantum computation.

$\mathbb{Z}_N$ parafermion zero modes \cite{FendleyJSM12,AliceaARCMP16} are generalizations of Majorana fermions (MFs) which correspond to the case $N=2$.
There have been several proposals to realize parafermions using quantum Hall states \cite{BarkeshliPRX12,LindnerPRX12,ChengPRB12,ClarkeNatComm13,VaeziPRB13,BarkeshliPRX14} and  coupled wires \cite{OregPRB14,KlinovajaPRB14,KlinovajaPRL14} all of which involve strong electron-electron interactions in some form.
In addition to the proposals in condensed matter settings, twist defects \cite{BombinPRL10} introduced in the $\mathbb{Z}_N$ toric code model have a quantum dimension of $\sqrt{N}$ and behave as parafermions for general $N$ \cite{YouPRB12,BrownPRL13,ZhengPRB15}.
Compared to Majorana fermions, parafermions provide a denser set of qudit rotations.
Recently, a 2D lattice model of $\mathbb{Z}_3$ parafermions is shown to support more exotic Fibonacci anyons \cite{StoudenmirePRB15} which enable the universal set of gates.
Therefore, parafermions are computationally more powerful and could potentially lead to the solution for universal topological quantum computation \cite{BarkeshliArXiv15}.

The exciting progress of experiments and theoretical ideas around Majorana fermions and parafermions motivates us to go beyond how to search for those exotic anyons and look into how to manipulate them for the purpose of 1) demonstrating non-Abelian exchange statistics as a short-term mission and 2) implementing quantum algorithms with topological quantum gates as a long-term goal.
Traditional approach of TQC involves encoding in the topological protected subspace, initialization and readout of the logical qubits/qudits via measurements of topological charges, and braiding operations to implement the quantum gates by slowly moving anyons around each other.
Braiding has to be slow enough compared to the time scale set by the energy gap to avoid exciting quasiparticles, and fast enough compared to the time scale set by the residual energy splitting between topological degenerate states \cite{KnappPRX16}. 
This makes braiding a challenging task.

Alternatively, braiding can be done using either interaction-based proposals \cite{vanHeckNJP12,BurrelloPRA13} or a measurement-only approach \cite{BondersonPRL08, WoottonJPA05, HutterPRX15} without physically moving anyons.
These two approaches are shown to be equivalent \cite{BondersonPRB13} and they can avoid diabatic errors associated with moving anyons. In particular, measurement-only approach will allow us to concentrate on improving measurement fidelity alone because that is the only type of operation required for MOTQC in order to initialize, braid and readout topological qubits/qudits.
However, the MOTQC proposal uses a series of topological charge forced measurements to simulate braiding.
Each forced measurement is a probabilistic ``repeat-until-success'' process, and hence has an inherent uncertainty with respect to the number of operations. 
This poses a challenge to synchronize the clock for computations running in parallel.

In this paper, we propose a forced-measurement-free measurement-based braiding (FMF-MBB) protocol.
Each braiding exchange can be simulated by performing three measurements supplemented by a correction operator.
Using the examples of Majorana fermions and parafermions, we show that the correction operator compensates for charge transfers among anyons occurred during the three measurements.
Furthermore, we demonstrate that we can apply the correction operator in software which greatly simplifies the protocol.
We show explicitly how to apply the FMF-MBB protocol to the demonstration of braiding statistics and measurement-only TQC.

\section{Measurement-based braiding without forced measurements}
\subsection{Diagrammatic Representation}
Following Refs.\,\cite{KitaevAP06,BondersonThesis,BondersonAP08,BondersonPRL08,BondersonPRB13,BarkeshliArXiv14}, we employ a diagrammatic representation of anyonic states and operators to describe a general anyon model.
This representation encapsulates the topological properties of anyons independent of specific physical model.
In general, an anyon model is defined by 1) a set $\mathcal{C}$ of topological charges $\mathit{a,b,c},\dots\in \mathcal{C}$ carried by anyons, 2) fusion rules specifying how topological charges are splitted or combined, and 3) braiding rules specifying what happens to the anyonic state once two anyons exchange positions.
There is a unique vacuum charge $I$ which has trivial fusion and braiding rules.
For each charge $a$, there exists a unique conjugate charge $\bar{a}$ which can be generated from vacuum $I$ together with $a$.
The associative fusion algebra defines the fusion rules as
\begin{equation}
 a\times b=\sum_{\mathit{c}\in\mathcal{C}}N^c_{ab}c
 \label{eq:fusion_rules}
\end{equation}
where the fusion multiplicity $N^{c}_{ab}$ specifies the number of possible ways for charges $a$ and $b$ to fuse into charge $c$.
The associated fusion and splitting Hilbert spaces $V^c_{ab}$ and $V^{ab}_c$ have the same dimension of $N^c_{ab}$.
The states in fusion and splitting spaces can be represented by trivalent vertices
\begin{eqnarray}
 (d_c/d_ad_b)^{1/4} \raisebox{-4.0mm}{ \includegraphics[scale=0.3]{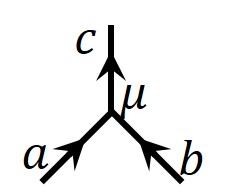} } &=&\langle a,b;c,\mu| \in V^c_{ab}, \\
 (d_c/d_ad_b)^{1/4} \raisebox{-5.3mm}{ \includegraphics[scale=0.3]{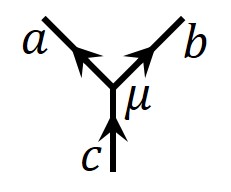} }&=&| a,b;c,\mu\rangle \in V^{ab}_c,
\end{eqnarray}
where $d_a$ is the quantum dimension of charge $\mathit{a}$ and $\mu=1,2,\dots,N^c_{ab}$ is the vertex label of basis.
Most anyon models of physical interest and which are the ones we will consider have no fusion multiplicity, i.e., $N^c_{ab}=0$ or $1$, and therefore we will leave the vertex label $\mu$ implicit hereafter.
The state space involving more than one fusion or splitting obeys associativity determined by $F$-moves
\begin{eqnarray}
 \raisebox{-5.0mm}{ \includegraphics[scale=0.3]{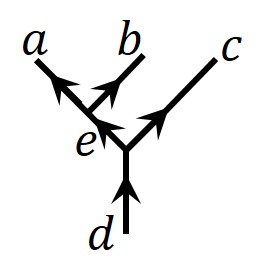} } &=&  \sum_f \left[F^{abc}_d\right]_{ef} \raisebox{-5.0mm}{ \includegraphics[scale=0.3]{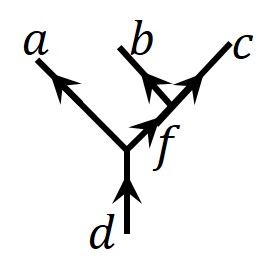} }, \label{eq:Fmoves-1}\\
  \raisebox{-5.0mm}{ \includegraphics[scale=0.3]{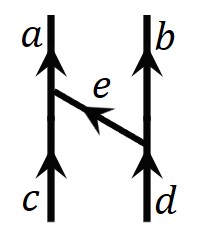} }&=& 
\sum_f\left[F^{ab}_{cd}\right]_{ef} \raisebox{-8.0mm}{ \includegraphics[scale=0.3]{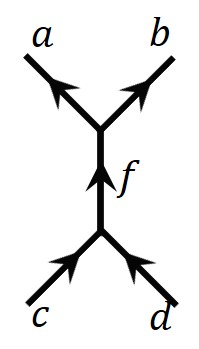} } , \label{eq:Fmoves-2}
\end{eqnarray}
where $\left[F^{ab}_{cd}\right]_{ef}=\sqrt{\frac{d_ed_f}{d_ad_d}}\left[F^{ceb}_f\right]^{*}_{ad}$.
The counter-clockwise braiding operator can be represented diagrammatically
\begin{equation}
 R_{ab}= \raisebox{-4.0mm}{ \includegraphics[scale=0.35]{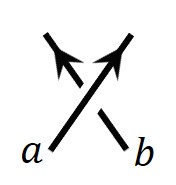} } = \sum_c \sqrt{\frac{d_c}{d_ad_b}}R^{ab}_c \raisebox{-7.0mm}{ \includegraphics[scale=0.3]{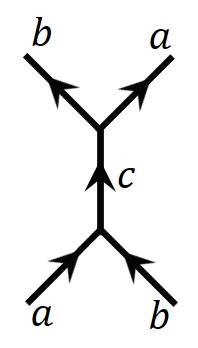} }
\label{eq:braiding}
 \end{equation}
where $R^{ab}_c$ is the phase acquired after exchanging charges $a$ and $b$ which fuse together into charge $c$. 

In this paper, we consider projective measurements of topological charges and physical examples include interferometry measurements \cite{DasSarmaPRL05,SternPRL06,BondersonPRL06,BondersonAP08,NayakRMP08}, topological blockade readout of topological charge \cite{vanHeckPRL13}, magnetic flux controlled parity readout using a top-transmon system \cite{KnappPRX16}, and electric charge sensing using a quantum point contact or a quantum dot \cite{AasenPRX16}.
In the diagrammatic representation, the projector of two anyons with charges $a_1$ and $a_2$ projected onto collective charge $b_{12}$ is given by
\begin{equation}
 \Pi^{(12)}_{b_{12}}=\sqrt{\frac{d_{b_{12}}}{d_{a_1}d_{a_2}}}  \raisebox{-7.0mm}{ \includegraphics[scale=0.3]{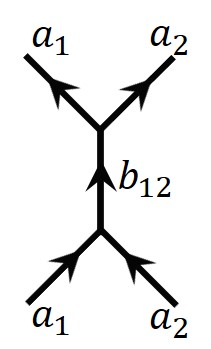} }.
\label{eq:projector}
 \end{equation}

\subsection{FMF-MBB Protocol}

\begin{figure}[tb!]
\centering
\includegraphics[width=0.45\textwidth]{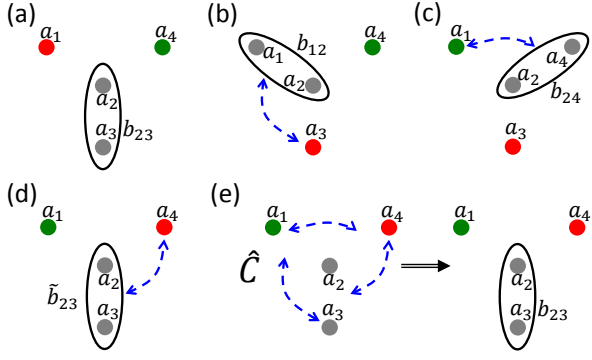}
\caption{(color online) FMF-MBB protocol to exchange anyons $a_1$ and $a_4$. (a) Initialization of ancilla anyons $a_2$ and $a_3$ into collective charge $b_{23}$. (b)-(d) Projective measurements of $a_1$ and $a_2$ into charge $b_{12}$, $a_2$ and $a_4$ into $b_{24}$, and $a_2$ and $a_3$ into $\tilde{b}_{23}$. The blue dash lines indicates charge transfers during the measurements. (e) The correction operator $\hat{C}$ is applied to undo all the charge transfers.}
\label{fig:Fig1-FMF}
\end{figure}

The key insight of the MOTQC protocol is to teleport anyonic state via projective measurements \cite{BondersonPRL08}.
As shown in Fig.\,\ref{fig:Fig1-FMF}(a), one can first initialize the ancilla anyons $a_2$ and $a_3$ into the collective charge $b_{23}$.
Then, one can perform a projective measurement of the collective charge $b_{12}$ of $a_1$ and $a_2$ [Fig.\,\ref{fig:Fig1-FMF}(b)].
It is apparent that anyonic state encoded in $a_1$ is teleported to $a_3$ if $b_{12}=b_{23}$.
In the case that $b_{12}\ne b_{23}$, one can go back to the initialization step and then measure $b_{23}$ and then $b_{12}$ again. Such a process can be repeated until the desired outcome $b_{12}=b_{23}$ is obtained.
This is the so-called forced measurement \cite{BondersonPRL08}.

In general, without imposing forced measurements, at each step in Fig.\,\ref{fig:Fig1-FMF}(b)-(d) there will be charge transfers between anyons \cite{BondersonPRB13,KnappPRX16} in addition to the anyonic state teleportation.
This is the essence of our FMF-MBB protocol:~we accept and keep track of the measurement results we obtain in Fig.\,\ref{fig:Fig1-FMF}(b)-(d), and supplement the final state with a correction operation to undo the charge transfers as shown in Fig.\,\ref{fig:Fig1-FMF}(e).
The resulting state is the same as the initial state in Fig.\,\ref{fig:Fig1-FMF}(a) except that $a_1$ and $a_4$ are exchanged.
Using Eq.\,(\ref{eq:projector}), we can write down the product of three-projective-measurement (TPM) operator $\hat{M}_{14,23}$ describing the projectors in Fig.\,\ref{fig:Fig1-FMF}(b)-(d) and $\Pi^{(23)}_{b_{23}}$
\begin{eqnarray}
&&\hat{M}_{14,23} \Pi^{(23)}_{b_{23}} = \Pi^{(23)}_{\tilde{b}_{23}} \Pi^{(24)}_{b_{24}} \Pi^{(12)}_{b_{12}} \Pi^{(23)}_{b_{23}} \nonumber \\
&&   = \mathcal{N}_1\raisebox{-25.0mm}{ \includegraphics[scale=0.3]{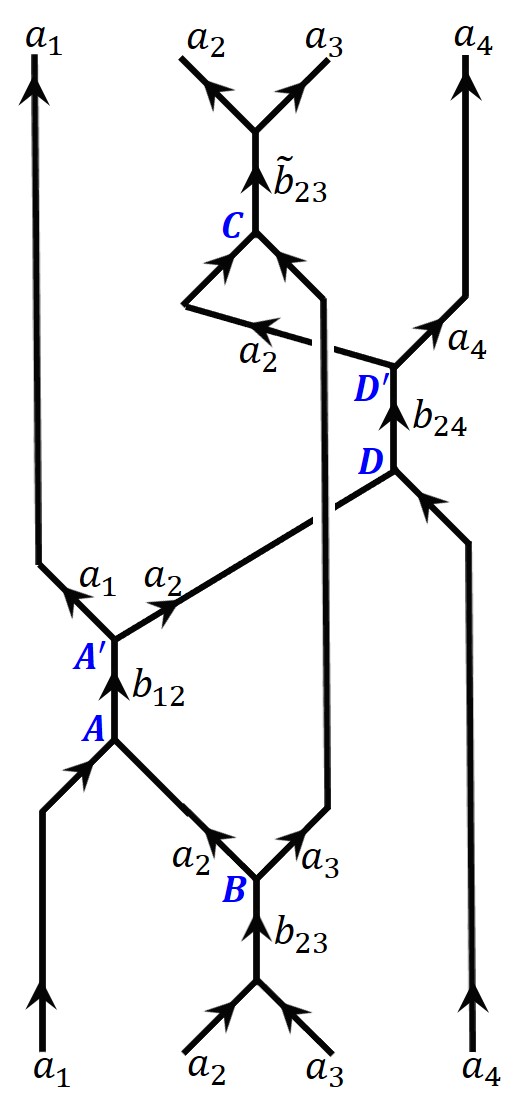} }=\mathcal{N}_1\raisebox{-11.0mm}{ \includegraphics[scale=0.3]{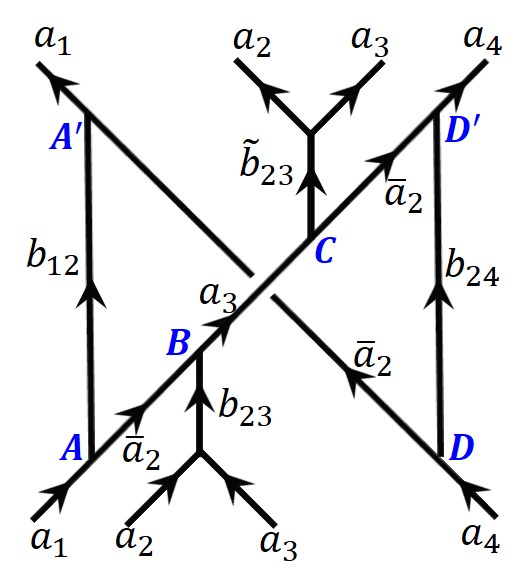} },
 \label{eq:Projectors-1}
\end{eqnarray}
where $\mathcal{N}_1$ is the normalization factor. Note that $\Pi^{(23)}_{b_{23}}$ is not part of the TPM operator.

To further proceed with the calculation, it is essential to restrict ourselves to the case that \textit{the intermediate collective charges $b_{23}$, $b_{12}$, $b_{24}$ and $\tilde{b}_{23}$ are Abelian}; otherwise the resulting operation will not be unitary and hence can not simulate braiding exchange \cite{BondersonPRB13}. 
This includes the Ising and $\mathbb{Z}_N$ parafermion models but not the Fibonacci anyon model.
Using the identities defined by $F-$moves in Eqs.\,(\ref{eq:Fmoves-1})-(\ref{eq:Fmoves-2}) and the Abelian-ness of $b_{12}$ and $b_{24}$, we can move $A^{\prime}$ and $D^{\prime}$ to merge with $D$ and $A$ respectively, and there is only an additional phase factor
\begin{equation}
 \hat{M}_{14,23} \Pi^{(23)}_{b_{23}}=\mathcal{N}_2 \raisebox{-11.0mm}{ \includegraphics[scale=0.3]{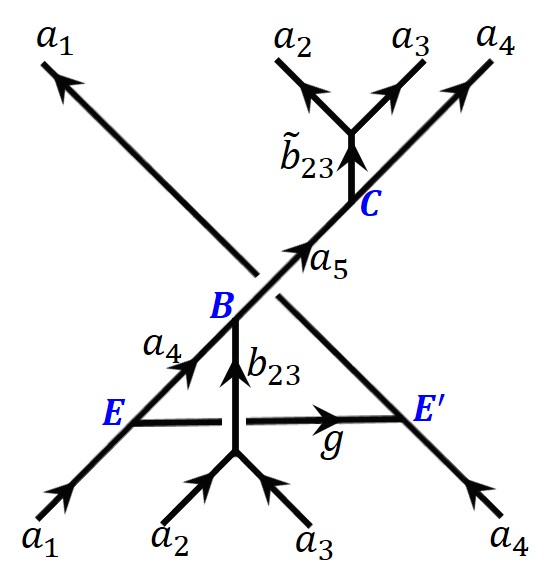} },
 \label{eq:Projectors-2}
\end{equation}
where $g=b_{12}\times \bar{b}_{24}$ and $a_5=a_4\times a_2\times a_3$.
Using Eq.\,(\ref{eq:Fmoves-2}), we can rewrite Eq.\,(\ref{eq:Projectors-2}) as
\begin{equation}
 \hat{M}_{14,23} \Pi^{(23)}_{b_{23}}=\mathcal{N}_3 \raisebox{-15.0mm}{ \includegraphics[scale=0.3]{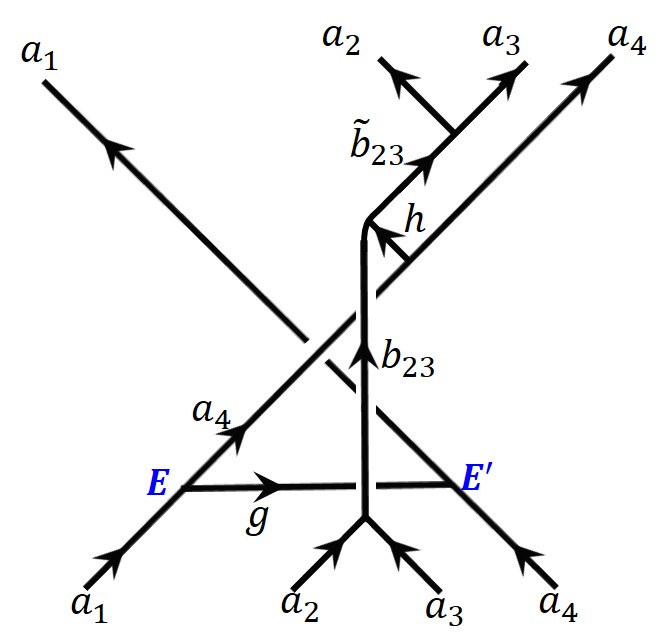} },
 \label{eq:Projectors-3}
\end{equation}
where $h=\tilde{b}_{23}\times \bar{b}_{23}$. Eq.\,(\ref{eq:Projectors-3}) provides a clear physical picture: compared to the forced measurement protocol, there are also charge transfers during the measurements in Fig.\,\ref{fig:Fig1-FMF}(b)-(d).
These charge transfers are encoded in two processes in Eq.\,(\ref{eq:Projectors-3}), namely exchange of Abelian charge $g$ between $a_1$ and $a_4$, and exchange of charge $h$ between ($a_2$, $a_3$) and ($a_1$, $a_4$).

After further manipulations utilizing the identities in Eqs.(\ref{eq:Fmoves-1})-(\ref{eq:braiding}), $\hat{M}_{14,23}$ is brought into a form close to the desired braiding-exchanged result
\begin{eqnarray} 
&& \hat{M}_{14,23} \Pi^{(23)}_{b_{23}} \nonumber \\
&& =\mathcal{N}_4 \sum_{cd}\left[F^{a_3a_4}_{a_3a_4}\right]_{hd} \left[F^{a_4ga_4}_{c}\right]_{a_1a_1} R^{a_4a_1}_{c} \raisebox{-12.0mm}{ \includegraphics[scale=0.3]{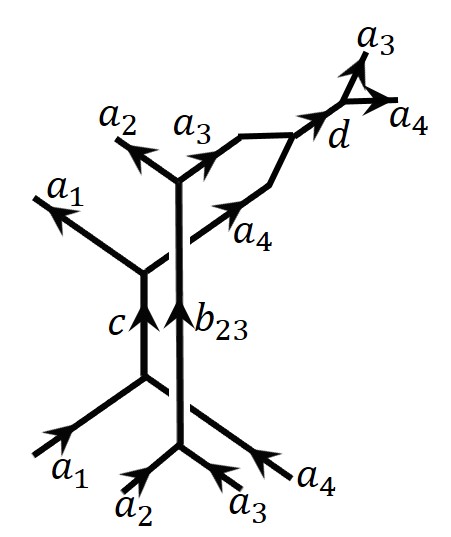} } \nonumber \\
&& = \mathcal{N}_4 \sum_{cd}\left[F^{a_3a_4}_{a_3a_4}\right]_{hd} \left[F^{a_4ga_4}_{c}\right]_{a_1a_1} R^{a_4a_1}_{c}  \Pi^{(34)}_{d}  \Pi^{(14)}_{c}  \Pi^{(23)}_{b_{23}}. \nonumber \\
 \label{eq:Projectors-4}
\end{eqnarray}
Eq.\,(\ref{eq:Projectors-4}) is applicable to models of either anyons or defects \cite{BarkeshliArXiv14} which exhibit interesting projective non-Abelian statistics as long as the collective charges $b_{23}$, $b_{12}$, $b_{24}$ and $\tilde{b}_{23}$ are Abelian.

\subsection{Effective Braiding for Parafermions}

To dis-entangle the effective braiding from the charge transfers, we apply the result to the $\mathbb{Z}_N$ parafermion model \cite{FendleyJSM12,AliceaARCMP16} which is physically relevant  ($N=2$ is the Majorana fermion case).
For $\mathbb{Z}_N$ parafermions, there are $N$ fusion states $|q\rangle$ of two parafermions $\sigma$, where $q=0,\dots,N-1$ is the Abelian charge (defined modulo $N$). 
The $F$-matrix of parafermions is given by \cite{VaeziPRB13,BarkeshliPRB13,BarkeshliArXiv14}
\begin{equation}
 \left[F^{\sigma a \sigma}_{b}\right]_{\sigma\sigma}=\omega^{-ab},\;\;\omega=\text{e}^{\frac{i2\pi}{N}}
\label{eq:FMatrice-PF}
 \end{equation}
where $a$ and $b$ are Abelian charges. 
Plugging Eq.\,(\ref{eq:FMatrice-PF}) into Eq.\,(\ref{eq:Projectors-4}), we have
\begin{flalign}
 \hat{M}_{14,23} \Pi^{(23)}_{b_{23}}=\text{e}^{i\phi} \sum_{cd} \omega^{h d-g c}   R^{\sigma_4\sigma_1}_{c}  \Pi^{(34)}_{d}  \Pi^{(14)}_{c}  \Pi^{(23)}_{b_{23}}, 
\label{eq:Projectors-5}
 \end{flalign}
where $\phi$ is an overall phase. 
In addition, the braiding and parity operators of parafermion modes $\gamma_i$ and $\gamma_j$ act on the fusion states in the following ways
\begin{eqnarray}
 &&\hat{R}_{ij}|q\rangle_{ij}=R^{\sigma_i\sigma_j}_q|q\rangle_{ij},\;\;\hat{P}_{ij}|q\rangle_{ij}=\omega^q |q\rangle_{ij},\nonumber \\
 &&\hat{P}_{ij}=\omega^{\frac{N+1}{2}}\gamma^{\dagger}_i\gamma_j.
\label{eq:BP-PF}
 \end{eqnarray}
Using Eq.\,(\ref{eq:BP-PF}), we can express Eq.\,(\ref{eq:Projectors-5}) in operator form
\begin{equation}
 \hat{M}_{14,23} \Pi^{(23)}_{b_{23}}=\text{e}^{i\phi} \sum_{cd} (\hat{P}_{34})^{h} \Pi^{(34)}_{d}  (\hat{P}_{14})^{-g} \hat{R}_{14} \Pi^{(14)}_{c}  \Pi^{(23)}_{b_{23}}.
 \label{eq:Projectors-6}
\end{equation}
Now, we supplement a correction operator $\hat{C}_{14,23}^{g,h}$
\begin{eqnarray}
 \hat{C}_{14,23}^{g,h} [\hat{M}_{14,23} \Pi^{(23)}_{b_{23}}] &=& \text{e}^{i\phi} \hat{R}_{14} \Pi^{(23)}_{b_{23}},\nonumber \\
  \hat{C}_{14,23}^{g,h}&=& (\hat{P}_{14})^{g}(\hat{P}_{34})^{-h}.
\label{eq:Projectors-7}
 \end{eqnarray}
It becomes apparent that the ancilla parafermions $2$ and $3$ return to their initialized collective charge state $b_{23}$ and an effective braiding exchange is achieved between parafermions $1$ and $4$ [Fig.\,\ref{fig:Fig1-FMF}(e)]
\begin{equation}
 \hat{C}_{14,23}^{g,h} \hat{M}_{14,23} = \text{e}^{i\phi} \hat{R}_{14}.
\label{eq:Projectors-8}
 \end{equation}
Eq.\,(\ref{eq:Projectors-8}) is the central result of our paper and the key insight here is to undo the charge transfers occurred during the measurements to realize the desired braiding operation.
Taking $N=2$, Eq.\,(\ref{eq:Projectors-7}) is consistent with our previous results of Majorana fermions \cite{ZhengPRB15} obtained using a wave-function approach.

\section{How to apply the correction operator $\hat{C}$?}
\subsection{Hardware-Implemented Correction Operator}

In principle, it is possible to apply the correction operator $\hat{C}$ in a topologically protected way on the hardware.
For the special case of Majorana fermions, one possible approach is to use the Aharonov-Casher (AC) effect \cite{AharonovPRL84,FriedmanPRL02}.
The parity operator $\hat{P}_{ij}$ is the Pauli $\hat{\sigma}_z$ operator in the logical space of MFs $\gamma_i$ and $\gamma_j$.
Applying $\hat{P}_{ij}$ imprints a $\pi$-phase difference between the logical state $|0\rangle_{ij}$ and $|1\rangle_{ij}$.
This can be achieved using the setup for the interferometry experiments proposed by Clarke and Shtengel \cite{ClarkePRB10}, and Grosfeld and Stern \cite{GrosfeldPNAS11}.
The Majorana modes are hosted by the inner superconducting island using the Majorana wire approach \cite{LutchynPRL10,OregPRL10}.
A Josephson vortex (fluxon) can be generated on demand in the circular Josephson junction and driven to circulate the loop by an applied supercurrent \cite{UstinovAPL02,FriedmanPRL02,WallraffNat03,BeckPRL05}.
Looping the fluxon around the inner superconducting island once will lead to a phase $\pi Q/2\mathrm{e}$ due to the AC effect \cite{AharonovPRL84,FriedmanPRL02,HasslerNJP10}, where $Q$ is the charge on the island.
Therefore, an exact $\pi$-phase difference will be produced between $|0\rangle_{ij}$ and $|1\rangle_{ij}$ states, which is equivalent to applying the parity operator.

\subsection{Software-Implemented Correction Operator}
For quantum computation with $\mathbb{Z}_N$ parafermions, we show that the correction operator can be applied in software to completely avoid potential errors introduced by the hardware approach, similar to how Pauli operations can be implemented in surface codes \cite{FowlerPRA12}. In particular, we consider the implementation of correction operators in software in three cases.
Since braiding is the building block of topological quantum computation, we first show the case of how to implement correction operator in software for a set of two braidings. Next, we study the case of implementing correction operator in software for a braiding operation followed by a charge measurement. Those two cases form the building blocks for the third case of applying in software correction operators in a generic computation comprising Clifford gates and charge measurements. 
\subsubsection{Case A: two braidings}
\label{caseA}

Suppose we want to implement a braiding operation $\hat{R}_{i_1i_4}$ to exchange computational parafermions $i_1$ and $i_4$ using ancilla parafermions $i_2$ and $i_3$, which is followed by the braiding exchange $\hat{R}_{j_1j_4}$ of computational parafermions $j_1$ and $j_4$ using ancilla $j_2$ and $j_3$. We consider general braidings, and hence computational parafermion $i_1$ or $i_4$ can be the same as $j_1$ or $j_4$. The same holds for ancilla parafermions $i_2$, $i_3$, $j_2$ and $j_3$. However, ancilla parafermions are solely used for assisting braiding and are different from computational parafermions.
According to Eq.\,(\ref{eq:Projectors-8}), these two braidings can be simulated by FMF-MBB protocol
\begin{equation}
 \hat{R}_{j_1j_4}\hat{R}_{i_1i_4}=[\hat{C}_{j_1j_4,j_2j_3}^{g_j,h_j} \hat{M}_{j_1j_4,j_2j_3}][\hat{C}_{i_1i_4,i_2i_3}^{g_i,h_i} \hat{M}_{i_1i_4,i_2i_3}],
\label{eq:TwoBrainding_before}
 \end{equation}
where 
\begin{equation}
g_k=b_{k_1k_2}\times \bar{b}_{k_2k_4},\; h_k=\tilde{b}_{k_2k_3}\times \bar{b}_{k_2k_3}, k=i, j, 
\label{eq:gk_definition} 
\end{equation}
keep track of intermediate charge transfers. 
Hereafter, we neglect the trivial overall phase factor.
The TPM operator for parafermions is given by
\begin{equation}
\hat{M}_{k_1k_4,k_2k_3}  = \Pi^{(k_2k_3)}_{\tilde{b}_{k_2k_3}} \Pi^{(k_2k_4)}_{b_{k_2k_4}} \Pi^{(k_1k_2)}_{b_{k_1k_2}} , \\
\end{equation}
where the projective charge measurement of parafermions can be expressed in terms of the parity operator
\begin{equation}
\Pi^{(pq)}_{b_{pq}}=\sum^N_{\ell=1}(\omega^{-b_{pq}}\hat{P}_{pq})^{\ell}.
\end{equation}
Parafermions obey the commutation relation 
\begin{equation}
\gamma_p\gamma_q=\omega^{\text{sgn}(q-p)}\gamma_q\gamma_p,
\label{eq:comm_parafermion}
\end{equation}
where $\text{sgn}(q-p)$ shows the importance of the relative ordering of $\gamma_q$ and $\gamma_p$ when $\omega^{-1}\ne \omega$.

Without loss of generality, we assume $j_{1-4}\le i_{1-4}$ in the relative ordering of parafermion modes \cite{FendleyJSM12}.
In this case, we can commute $\hat{C}_{i_1i_4,i_2i_3}^{g_i,h_i}$ through $\hat{M}_{j_1j_4,j_2j_3}$ using Eq.\,(\ref{eq:comm_parafermion})
\begin{equation}
 \hat{R}_{j_1j_4}\hat{R}_{i_1i_4}=\hat{C}_{j_1j_4,j_2j_3}^{g_j,h_j} \hat{C}_{i_1i_4,i_2i_3}^{g_i,h_i} \hat{M}^{\prime}_{j_1j_4,j_2j_3} \hat{M}_{i_1i_4,i_2i_3},
\label{eq:TwoBrainding_after_para}
 \end{equation}
where
\begin{eqnarray}
&& \hat{M}^{\prime}_{j_1j_4,j_2j_3}= \Pi^{(j_2j_3)}_{\tilde{b}^{\prime}_{j_2j_3}} \Pi^{(j_2j_4)}_{b^{\prime}_{j_2j_4}} \Pi^{(j_1j_2)}_{b^{\prime}_{j_1j_2}}, \label{eq:O_updated_para}\\
&& b^{\prime}_{j_1j_2} = b_{j_1j_2} + [\delta_{j_1i_1}-\delta_{j_1i_4}]g_i+[\delta_{j_1i_4}+\delta_{j_2i_3}]h_i, \\
&& b^{\prime}_{j_2j_4} = b_{j_2j_4} + [\delta_{j_4i_4}-\delta_{j_4i_1}]g_i-[\delta_{j_4i_4}+\delta_{j_2i_3}]h_i, \\
&& \tilde{b}^{\prime}_{j_2j_3} = \tilde{b}_{j_2j_3} +[\delta_{j_3i_3}-\delta_{j_2i_3}]h_i.
\label{eq:s_updated_para}
\end{eqnarray}
Here, $g_i=b_{i_1i_2}\times \bar{b}_{i_2i_4}$ and $h_i=\tilde{b}_{i_2i_3}\times \bar{b}_{i_2i_3}$ are the charge transfers between parafermions $i_1$-$i_4$ expressed in terms of the charge readouts. To simplify the above expression, we introduce the short-hand notations
\begin{eqnarray}
&& R_k\equiv \hat{R}_{k_1k_4}, C_k \equiv \hat{C}_{k_1k_4,k_2k_3}^{g_k,h_k}, M_k\equiv \hat{M}_{k_1k_4,k_2k_3},\\
&& \mathbf{b_j} = \begin{bmatrix}
         b_{j_1j_2} \\
         b_{j_2j_4} \\
         \tilde{b}_{j_2j_3}
        \end{bmatrix},   
     \mathbf{b^{\prime}_j} = \begin{bmatrix}
         b^{\prime}_{j_1j_2} \\
         b^{\prime}_{j_2j_4} \\
         \tilde{b}^{\prime}_{j_2j_3}
        \end{bmatrix}.
\end{eqnarray}
The intermediate charge measurement results are implicitly in the short-hand notations, and we can express Eq.\,(\ref{eq:O_updated_para})-Eq.\,(\ref{eq:s_updated_para}) as
\begin{eqnarray}
&& R_j R_i = (C_jM_j)(C_iM_i)=C_jC_iM^{\prime}_j M_i,  \\
&& M^{\prime}_j = \Pi^{(j_2j_3)}_{\tilde{b}^{\prime}_{j_2j_3}} \Pi^{(j_2j_4)}_{b^{\prime}_{j_2j_4}} \Pi^{(j_1j_2)}_{b^{\prime}_{j_1j_2}},\\
&& \mathbf{b^{\prime}_j}  = \mathbf{b_j} + \mathbf{\Delta_j(g_i,h_i) }, \label{eq:chargeupdate}
\end{eqnarray}
where
\begin{equation}
\mathbf{\Delta_j(g_i,h_i) } = \begin{bmatrix}
         (\delta_{j_1i_1}-\delta_{j_1i_4})g_i+(\delta_{j_1i_4}+\delta_{j_2i_3})h_i \\
         (\delta_{j_4i_4}-\delta_{j_4i_1})g_i-(\delta_{j_4i_4}+\delta_{j_2i_3})h_i \\
         (\delta_{j_3i_3}-\delta_{j_2i_3})h_i
        \end{bmatrix}
\label{eq:Delta}
\end{equation} 
is the difference between the charge measurement results before and after commuting $C_i$ through $M_j$. 
It is dependent on the intermediate measurement results encoded in $g_i$ and $h_i$ from the first TMP operator, and that whether the two set of parafermions are the same or not.
Therefore, the effect of postponing the correction operator is to have updated intermediate measurement results in the second TPM operator.

\subsubsection{Case B: a braiding followed by a charge measurement}
\label{caseB}

Next, we consider the case of a braiding operation $R_i$ followed by a charge measurement $\Pi_j$. $R_i$ realizes the braiding exchange between computational parafermions $i_1$ and $i_4$ assisted by the ancilla parafermions $i_2$ and $i_3$. According to Eq.\,(\ref{eq:Projectors-8}), $R_i$ can be simulated by a TPM operator $M_i$ followed by a correction operator $C_i$, where $M_i\equiv \hat{M}_{i_1i_4,i_2i_3}$ and $C_i\equiv \hat{C}_{i_1i_4,i_2i_3}^{g_i,h_i}$. The charge measurement $\Pi_j\equiv \Pi^{(j_1j_4)}_{c_{j_1j_4}}$ measures the collective charge $c_{j_1j_4}$ of computational parafermions $j_1$ and $j_4$, which can be the same as $i_1$ or $i_4$ but not $i_2$ and $i_3$. The operator for this case can be written as
\begin{equation}
\Pi_j R_i = \Pi_j (C_i M_i) = \Pi^{(j_1j_4)}_{c_{j_1j_4}} \hat{C}_{i_1i_4,i_2i_3}^{g_i,h_i} \hat{M}_{i_1i_4,i_2i_3}.
\end{equation}
Again, we assume $j_{1,4}\le i_{1-4}$ in the relative ordering of parafermion modes
After commuting the correction operator through the charge measurement operator using Eq.\,(\ref{eq:comm_parafermion}), we have
\begin{equation}
\Pi_j R_i = \Pi_j (C_i M_i) = C_i \Pi^{\prime}_j M_i,
\label{eq:Pi_1}
\end{equation}
where 
\begin{eqnarray}
&& \Pi^{\prime}_j  = \Pi^{(j_1j_4)}_{c^{\prime}_{j_1j_4}}, \label{eq:Pi_2} \\
&& c^{\prime}_{j_1j_4}=c_{j_1j_4} + \eta_j(g_i, h_i), \label{eq:Pi_3} \\
&& \eta_j(g_i, h_i) = (\delta_{j_1i_1}-\delta_{j_4i_1})g_i+(\delta_{j_4i_4}-\delta_{j_1i_4})(g_i-h_i), \;\;\;\;\;\; \label{eq:Pi_4}
\end{eqnarray}
encodes the change of the final measurement result due to commuting the correction operator through.

\begin{figure}[tb!]
\centering
\includegraphics[width=0.48\textwidth]{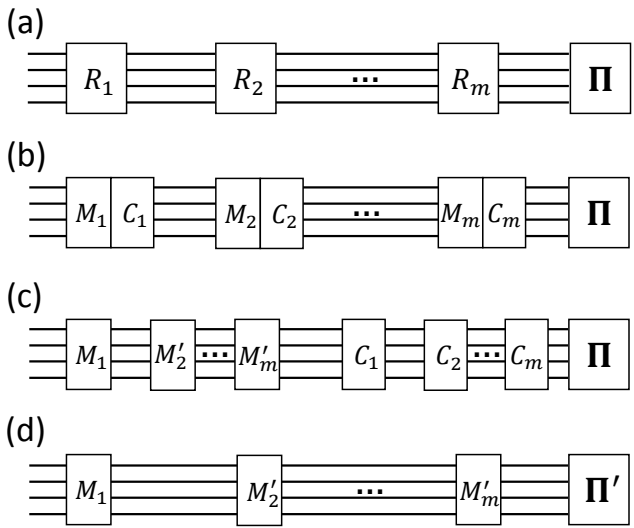}
\caption{Schematic of software-implemented correction operator for a generic computation using Clifford gates. 
(a) A series of $m$ braiding operations $R_1,\dots,R_m$ followed by $n$ projective measurements $\mathbf{\Pi}=(\Pi_1,\dots,\Pi_n)$. Here each line represents a computational parafermion and each braiding or measurement acts on two parafermions.
(b) Each braiding operation can be replaced by a TPM operator $M$ followed by a correction operator $C$ [see Eq.\,(\ref{eq:Projectors-8})].
(c) Commuting the correction operators $C_1,\dots,C_m$ through the TPM operators $M_2,\dots,M_m$. The TPM operators are updated to $M^{\prime}$. 
(d) Commuting the correction operators through the final measurements. The measurements are updated to $\mathbf{\Pi}^{\prime}$.}
\label{fig:Fig2-SICO}
\end{figure}

\subsubsection{Case C: general computation}
\label{caseC}
Now, for a general computation comprised of Clifford gates, we can break it down to a series of braiding operations $R_i,i=1,\dots,m$, followed by a set of final projective measurements $\mathbf{\Pi}=(\Pi_1, \Pi_2,\dots,\Pi_n)$ on the logical qudits to extract information as shown in Fig.\,\ref{fig:Fig2-SICO}(a) (we defer the discussion of non-Clifford gates to next section). Here, each braiding operation involves two computational parafermions and two ancilla parafermions, and each charge measurement involves two computational parafermions.
For simplicity, we keep the indices of the parafermions implicit in the definition of $R_i$ and $\mathbf{\Pi}$.

According to Eq.\,(\ref{eq:Projectors-8}), each braiding operation $R_i$ can be simulated by a combination of a TPM operator $M_i$ and a correction operator $C_i$  as shown in Fig.\,\ref{fig:Fig2-SICO}(b). According to the discussion in Sec.\,\ref{caseA}, we can commute $C_{m-1}$ through $M_m$, which then becomes $M^{\prime}_m$.
In general, for $M_i$ we need to commute $C_1,\dots,C_{i-1}$ through and $M_i$ becomes $M^{\prime}_i$ with a set of updated charge measurements
\begin{equation}
\mathbf{b^{\prime}_i}  = \mathbf{b_i} + \sum_{k=1}^{i-1}\mathbf{\Delta_i(g_k,h_k) },
\label{eq:b_update}
\end{equation}
where $\mathbf{b_i}$ is the set of charge measurements of $M_i$ and $\mathbf{\Delta_i(g_k,h_k)}$ is defined in Eq.\,(\ref{eq:chargeupdate}), encoding the change due to commuting $C_k$ through $M_i$ \footnote{The detailed form of $\mathbf{\Delta_i(g_k,h_k)}$ can be different from Eq.\,(\ref{eq:chargeupdate}) where a particular ordering of parafermions is assumed.}.
Fig.\,\ref{fig:Fig2-SICO}(c) illustrates the effect of commuting all the correction operators through the TPM operators.

In the final step of software-implemented correction operator, we commute all the correction operators through the final set of logic measurements $\mathbf{\Pi}$ following Sec.\,\ref{caseB}. According to Eq.\,(\ref{eq:Pi_1}) to Eq.\,(\ref{eq:Pi_4}), charge measurement $\Pi_j$ in $\mathbf{\Pi}$ will become $\Pi^{\prime}_j$ with updated charge measurement result
\begin{equation}
c^{\prime}_{j}=c_{j} + \sum^{m}_{k=1}\eta_j(g_k, h_k),
\label{eq:c_update}
\end{equation}
where $\eta_j(g_k, h_k)$ is defined in Eq.\,(\ref{eq:Pi_4}), encoding the change due to commuting $C_k$ through $\Pi_j$ \footnote{Again, the detailed form of $\eta_j(g_k, h_k)$ can be different from Eq.\,(\ref{eq:Pi_4}) where a particular ordering of parafermions is assumed.}. Once commuted through the final measurements, the correction operators have no impact on the computation any more. The final result of software-implemented correction operators is shown in Fig.\,\ref{fig:Fig2-SICO}(d).

Therefore, the general procedure for carrying out a computation in Fig.\,\ref{fig:Fig2-SICO}(a) is to follow the steps in Fig.\,\ref{fig:Fig2-SICO}(d) and record the charge measurement results $(\mathbf{b^{\prime}_1},\dots,\mathbf{b^{\prime}_m})$ of the TPM operators as well as the final measurements ($c^{\prime}_1, \dots, c^{\prime}_n$).
In order to extract the actual logic measurement results ($c_1,\dots,c_n$), we can do postprocessing in two steps. First, using Eq.\,(\ref{eq:b_update}) we can perform a backtracking to find $(\mathbf{b_1},\dots,\mathbf{b_m})$. For instance, $\mathbf{b_1}$ is the same as $\mathbf{b^{\prime}_1}$, which can be used to compute $\mathbf{\Delta_2(g_1,k_1)}$. Then, $\mathbf{b_2}$ can be calculated using $\mathbf{b^{\prime}_2}$ and $\mathbf{\Delta_2(g_1,k_1)}$. For general $i$, we use all the previously calculated $\mathbf{b_{1,\dots,i-1}}$ to compute $\mathbf{\Delta_i(g_k, h_k)}$ ($k=1,\dots,i-1$), which is further used to find $\mathbf{b_i}$. This way, we can calculate $(\mathbf{b_1},\dots,\mathbf{b_m})$ using $(\mathbf{b^{\prime}_1},\dots,\mathbf{b^{\prime}_m})$. The second step of postprocessing is to compute $\eta_j(g_k,h_k)$ using $(\mathbf{b_1},\dots,\mathbf{b_m})$. Then Eq.\,(\ref{eq:c_update}) will give us the desired actual measurements ($c_1,\dots,c_n$).
We summarize the postprocessing backtracking algorithm below. It runs in linear time in the number of braiding operations and hence can be carried out efficiently.
This is the essence of the software-assisted FMF measurement-based braiding.

\begin{algorithm}[H]
\caption{Postprocessing Backtracking Algorithm}
\label{alg::PBA}
\begin{algorithmic}[1]
\Require
$(\mathbf{b^{\prime}_1},\dots\mathbf{b^{\prime}_m},)$: intermediate charge measurements
$(c^{\prime}_1,\dots,c^{\prime}_n)$: measured charges in the final measurement of $\Pi^{\prime}$ in Fig.\,\ref{fig:Fig2-SICO}(d)
\Ensure
$(c_1,\dots,c_n)$: original measured charges in Fig.\,\ref{fig:Fig2-SICO}(a)
\State $\mathbf{b_1}=\mathbf{b^{\prime}_1} \rightarrow (g_1,k_1)$ [Eq.\,(\ref{eq:gk_definition})]
\For{$i=2$ to $m$}
\State $g_{i-1},k_{i-1} \rightarrow \mathbf{\Delta_i(g_{i-1},k_{i-1})}$ [Eq.\,(\ref{eq:Delta})]
\State $\mathbf{b^{\prime}_i}, \mathbf{\Delta_i(g_{1},k_{1})},\dots,\mathbf{\Delta_i(g_{i-1},k_{i-1})} \rightarrow \mathbf{b_i}$ [Eq.\,(\ref{eq:b_update})]
\State $\mathbf{b_i} \rightarrow g_i,k_i$ [Eq.\,(\ref{eq:gk_definition})]
\EndFor
\State $(g_1,h_1,\dots,g_m,h_m)\rightarrow \eta_j(g_i,h_i)$ [Eq.\,(\ref{eq:Pi_4})]
\For{$j=1$ to $n$}
\State $c^{\prime}_j,\eta_j(g_1,h_1),\dots,\eta_j(g_m,h_m)\rightarrow c_j$ [Eq.\,(\ref{eq:c_update})]
\EndFor
\end{algorithmic}
\end{algorithm}

 \begin{figure*}[tb]
\centering
\includegraphics[width=1.0\textwidth]{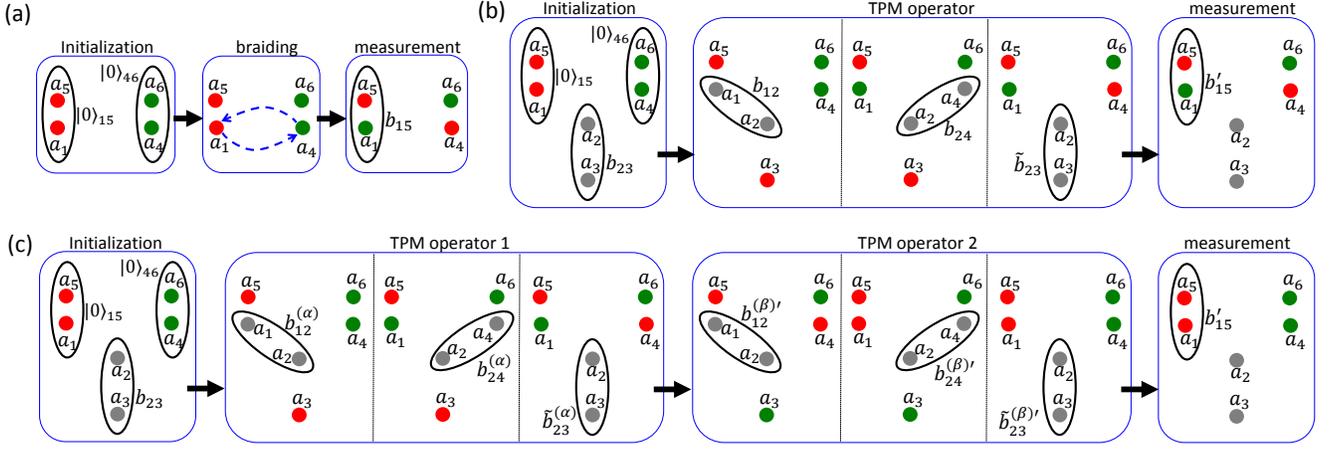}
\caption{(color online) Demonstration of braiding exchange statistics of Majorana fermions. (a) The usual way of demonstrating the non-Abelian statistics of Majorana fermions in three steps.
(b) FMF protocol for a single braiding exchange: the initialization and measurement steps are the same as (a), but the physical exchange of MFs is replaced by a set of three charge measurements. 
(c) FMF protocol for two braiding exchanges. The final charge readout $b^{\prime}_{15}$ is different from $b_{15}$ because the correction operator is applied in software. See Eq.\,(\ref{eq:parity-connection}) and Eq.\,(\ref{eq:TB-finalParity}) in the main text for the connection between $b_{15}$ and $b^{\prime}_{15}$.}
\label{fig:Fig2-FMF_braiding}
\end{figure*}

\section{Demonstration of non-Abelian statistics with FMF-MBB Protocol} 
Now, we show how to apply our software-assisted FMF-MBB protocol to the task of demonstrating non-Abelian statistics of Majorana fermions.
We do this by considering the two simplest examples of braiding experiments involving a single braiding exchange and two braiding exchanges.
\subsection{Single-Braiding Experiment}
Fig.\,\ref{fig:Fig2-FMF_braiding}(a) shows the usual way of using a single braiding exchange to demonstrate non-Abelian statistics of Majorana fermions in three steps.
First, Majorana pairs of ($1$,$5$) and ($4$,$6$) are initialized in the even parity states.
Then MFs $1$ and $4$ are moved physically to exchange their positions \cite{AasenPRX16}.
Finally, the fermion charge $b_{15}$ of the pair ($1$,$5$) is measured.
Ideally, the parity of $b_{15}$ has $50\%$ probability to be even or odd, which is a direct evidence of the non-Abelian statistics of Majorana fermions.

Fig.\,\ref{fig:Fig2-FMF_braiding}(b) shows the approach of FMF-MBB using ancilla Majorana fermions $a_2$ and $a_3$.
We replace the physical exchange of MFs $1$ and $4$ by the software-assisted FMF-MBB which is followed by a final charge measurement of the pair ($1$, $5$).
The parity of $b^{\prime}_{15}$ can be different from that of $b_{15}$. The operator describing the usual braiding in Fig.\,\ref{fig:Fig2-FMF_braiding}(a) is given by
\begin{equation}
 \hat{F}_1 = \Pi^{(15)}_{b_{15}} \hat{R}_{14}.
 \label{eq:F1}
\end{equation}
Using Eq.\,(\ref{eq:Projectors-8}), we can replace the braiding operator by the set of three measurements shown in Fig.\,\ref{fig:Fig2-FMF_braiding}(b) followed by the correction operator (ignoring the trivial phase factor)
\begin{equation}
 \hat{F}_1 = \Pi^{(15)}_{b_{15}} \hat{C}_{14,23}^{g,h} \hat{M}_{14,23},
\label{eq:F2}
 \end{equation}
where
\begin{eqnarray}
 g&=&b_{12}\otimes \bar{b}_{24},\\
 h&=&\tilde{b}_{23} \otimes \bar{b}_{23}. 
\end{eqnarray}
After commuting the correction operator through the final charge measurement, we have
\begin{equation}
 \hat{F}_1 = \hat{C}_{14,23}^{g,h}  \Pi^{(15)}_{b^{\prime}_{15}} \hat{M}_{14,23},
\label{eq:F3}
 \end{equation}
where 
\begin{equation}
 b^{\prime}_{15} = b_{15} + g.
 \label{eq:parity-connection}
\end{equation}
Therefore, for the single braiding experiment, one just needs to follow the steps depicted in Fig.\,\ref{fig:Fig2-FMF_braiding}(b) and the actual charge $b_{15}$ is connected to the measured charged $b^{\prime}_{15}$ via Eq.\,(\ref{eq:parity-connection}).

\subsection{Double-Braiding Experiment}
The case of two braidings can be approached in a similar way as shown in Fig.\,\ref{fig:Fig2-FMF_braiding}(c).
The operator describing the action of two braiding exchanges followed by a charge readout is given by
\begin{equation}
 \hat{F}_2 = \Pi^{(15)}_{b_{15}} \hat{R}^{(\beta)}_{14} \hat{R}^{(\alpha)}_{14}.
\label{eq:initial-F2}
 \end{equation}
Using Eq.\,(\ref{eq:Projectors-8}), we have (again, neglecting the overall phase factor)
\begin{equation}
 \hat{F}_2 = \Pi^{(15)}_{b_{15}} [\hat{C}_{14,23}^{g_{\beta},h_{\beta}} \hat{M}^{(\beta)}_{14,23}] [\hat{C}_{14,23}^{g_{\alpha},h_{\alpha}} \hat{M}^{(\alpha)}_{14,23}],
\end{equation}
where $\hat{M}^{(\alpha)}_{14,23}$ and $\hat{M}^{(\beta)}_{14,23}$ are the first and second TPM operators respectively and
\begin{eqnarray}
  g_{\alpha,\beta}&=&b^{(\alpha,\beta)}_{12} \otimes \bar{b}^{(\alpha,\beta)}_{24}, \label{eq:g_beta} \\
 h_{\alpha,\beta}&=&\tilde{b}^{(\alpha,\beta)}_{23}\otimes \bar{b}^{(\alpha,\beta)}_{23}. \label{eq:h_beta}
\end{eqnarray}
Finally, we can commute the two correction operators through the final charge measurement operator [Eq.\,(\ref{eq:TwoBrainding_after_para}) and Eq.\,(\ref{eq:F3})]
\begin{equation}
 \hat{F}_2 =  \hat{C}_{14,23}^{g_{\beta},h_{\beta}} \hat{C}_{14,23}^{g_{\alpha},h_{\alpha}} \Pi^{(15)}_{b^{\prime}_{15}}  \hat{M}^{(\beta)\prime}_{14,23}  \hat{M}^{(\alpha)}_{14,23},
\label{eq:final-F2}
 \end{equation}
where
\begin{eqnarray}
&& \hat{M}^{(\beta)\prime}_{14,23}= \Pi^{(23)}_{\tilde{b}^{(\beta)\prime}_{23}} \Pi^{(24)}_{b^{(\beta) \prime}_{24}} \Pi^{(12)}_{b^{(\beta)\prime}_{12}}, \\
&& b^{(\beta)\prime}_{12} = b^{(\beta)}_{12} + g_{\alpha}, \label{eq:TB-s1}\\
&& b^{(\beta)\prime}_{24} = b^{(\beta)}_{24} + g_{\alpha}-h_{\alpha}, \\
&& \tilde{b}^{(\beta)\prime}_{23} = \tilde{b}^{(\beta)}_{23} + h_{\alpha},\label{eq:TB-s3}\\
&& b^{\prime}_{15} = b_{15}  + g_{\alpha}+g_{\beta}. \label{eq:TB-finalParity}
\end{eqnarray}
In the two braiding experiment using our proposed FMF-MBB protocol, we have access to the charge readouts $b^{(\alpha)}$ from the first set of three measurements, $b^{(\beta)\prime}$ from the second set, and the final charge readout $b^{\prime}_{15}$.
We can first backtrack and calculate $b^{(\beta)}$ using Eqs.\,(\ref{eq:TB-s1})-(\ref{eq:TB-s3}), which yields $g_{\beta}$ and $h_{\beta}$ according to Eqs.\,(\ref{eq:g_beta})-(\ref{eq:h_beta}).
Then a further backtracking using Eq.\,(\ref{eq:TB-finalParity}) gives us the actual parity of $b_{15}$ after two braiding exchanges.
Here, the key is that the correction operator is a Pauli operator acting on the logical qubit made of two Majorana modes.
The TPM operator comprises three single-qubit projective measurements, which remain projective measurements with updated measurement results after commuting Pauli operators through.

Ideally, given an initially even parity state of ($1$,$5$), $b_{15}$ has a $100\%$ probability of flipping to the odd parity after two braidings as a direct consequence of the non-Abelian statistics of Majorana fermions. 
Here, the only experimental capability required is to perform quantum non-demolition (QND) readouts of the charge of Majorana pairs \cite{HasslerNJP10,AasenPRX16}.
Such a simplification will allow us not to worry about the diabatic errors associated with moving Majorana fermions and instead to concentrate on improving readout techniques.
In fact, because the measurements are QND, a natural way to boost the measurement fidelity is to simply repeat the same charge readout several times. 
In addition, FMF-MBB protocol removes the uncertainty associated with the number of measurements in the original MBB approach, and hence is more efficient experimentally.
Therefore, we believe our FMF-MBB approach can be an appealing avenue to the demonstration of non-Abelian braiding statistics.

\section{MOTQC with FMF-MBB Protocol}
Finally, we outline how to adapt our FMF protocol to the more ambitious long-term goal of measurement-only topological quantum computation using Majorana fermions and parafermions.
For Clifford gates, it is straightforward to apply the software-assisted FMF-MBB protocol because braiding itself is enough to carry out Clifford operations.

However, to complete the set of gates for universal quantum computation using Majorana fermions, one also needs to add the $\pi/8$-phase gate ($\hat{T}$ gate) which cannot be realized in a topologically protected way.
Fortunately, there exist protocols such as ``magic state distillation'' to generate a high fidelity $\hat{T}$ gate.
First, the distillation protocol starts with $15$ approximate copies of the ancilla state $|a\rangle=(|0\rangle+\text{e}^{i\pi/4}|1\rangle)/\sqrt{2}$.
For Majorana fermions, this can be done by initializing the state in $|0\rangle$ and performing a single-qubit $\pi/4$ rotation. One way to generate the single-qubit rotation is to bringing two Majorana modes together for a fixed amount of time such that the tunneling splitting imposes an approximate $\pi/4$ phase difference between $|0\rangle$ and $|1\rangle$.
Another possibility is to use hybrid systems that couple the Majorana fermions with other physical devices (e.g., tunnel junctions, flux qubits) to produce the desired ancillary states \cite{ClarkePRB10, HasslerNJP10,BondersonPRL10,JiangPRL11,BondersonPRL11}.
In the second step, these states are then projected onto the subspace of the Reed-Muller code \cite{RaussendorfAP06,BravyiPRA05}.
By subsequent stabilizer measurements, the states are encoded into a single logical qubit and it is a purified version of $|a\rangle$.
Finally, a $T$ gate can be applied to the data qubit after performing several Clifford operations and projective measurements on a small circuit made of the purified $|a\rangle$ state and the data qubit \footnote{See Fig.\,33 and Sec.\,XVI of Ref.\cite{FowlerPRA12} for a detailed description.}.

Notice that the first step to generate ancilla states is independent of how we carry out the braiding operations (no matter it is done via physical movements or our FMF-MBB protocol).
However, the second and third steps do heavily depend on the details of braiding since all the Clifford operations break down to sequences of braiding exchanges.
Fortunately, \textit{there are only two types of operations in the distillation protocol, namely Clifford operations and single-qubit projective measurements.}
Adopting the FMF-MBB protocol developed above, we can replace each braiding operator by the product of the TPM operator $\hat{M}$ and the correction operator $\hat{C}$ [Eq.\,(\ref{eq:Projectors-8})].
All the correction operators can then be commuted through the final measurements [Eq.\,(\ref{eq:final-F2})].
The single-qubit projective measurements remain Pauli measurements with updated measurement results dependent on the correction operators.
Therefore, the same software-assisted FMF-MBB procedure outlined for the single-braiding and two-braiding experiments applies equally well to the magic state distillation protocol and hence to MOTQC using Majorana fermions.

For $\mathbb{Z}_N$ parafermions, magic state distillation is well-studied for the case of prime $N$ \cite{CampbellPRX12}, and it was recently shown that all Clifford operations can be realized via braiding for odd $N$ \cite{HutterPRB16}.
Hence, all the above discussion carries over to the scenario of TQC using odd prime $N$ parafermions.

\section{Conclusion}
We have proposed a protocol of measurement-based braiding without forced measurements.
In particular, the braiding exchange is shown to be equivalent to a set of three measurements followed by a correction operation because we can always introduce correction operations to compensate the (topological) charge transfer during the measurement-based braiding.
Furthermore, for quantum computation with Majorana fermions or parafermions, we also show that the correction operator can be applied in software similar in spirit to how the Pauli operations can be implemented in surface codes.
Like the original MBB protocol, our FMF-MBB protocol removes the need for moving anyons physically and reduces the experimental requirement of braiding to the capability of performing projective measurements only.
Compared to the MBB protocol, it also removes the ambiguity in the number of measurements needed to realize a single braiding operation.
Finally, we show explicitly that such a simple braiding protocol can be applied to both the demonstration of non-Abelian braiding statistics and measurement-based topological quantum computation using Majorana fermions and parafermions.

\begin{acknowledgments}
We would like to thank Jukka Vayrynen, Aris Alexandradinata and Bernard van Heck for useful discussions.
We acknowledge support from ARL-CDQI, ARO, AFOSR MURI, the Alfred P. Sloan
Foundation, and the Packard Foundation.
\end{acknowledgments}

\bibliography{MOTQC}








\end{document}